# Decoherence and the Transactional Interpretation

R. E. Kastner
University of Maryland, College Park
31 March 2020

Abstract. This paper presents an analysis of decoherence resulting from the physically real non-unitarity, or 'objective reduction,' that occurs in the Transactional Interpretation (TI). Two distinct aspects of the decoherence process are identified and disambiguated; specifically, (i) the resolution of the basic measurement interaction with respect to the observable under study, and (ii) the effect on the measured system of repetition of the measurement interaction. It is shown that the measurement interaction as described in TI leads naturally to the same quantitative expression for the decoherence function as in the standard unitary-only account. However, unlike in the unitary-only approach, under TI, the reduced density operator for the measured system can legitimately be interpreted as representing the occurrence of an actual measurement result.

## 1. Introduction

Decoherence is a term used for a system's loss of interference between quantum states due to its interactions with external degrees of freedom. More formally, decoherence describes the vanishing of the off-diagonal elements of a system's density matrix with respect to a measurement 'pointer' basis, commonly position or some macroscopically observable parameter. The study of decoherence, pioneered by Joos, Zeh, Zurek, Omnès and others,[1] is a rigorous and well-developed research program that has amassed a large body of experimental corroboration. Thus, key aspects of the theoretical treatment of decoherence have demonstrated empirical validity.

However, the interpretation of the physical nature of the decohering system, in particular its relation to the measurement process, remains obscure in traditional accounts of decoherence. This is because the program was developed under the assumption that quantum theory involves only unitary (linear, deterministic) evolution. Under that approach, quantum correlations induced between the various interacting degrees of freedom—i.e., the system, its measuring device, the environment--persist indefinitely. It is assumed that the induced entanglement never ends, so that 'measurement' is no more than the establishment of such correlated entanglements. To obtain predictions about the measured system from the total density operator for all correlated systems, one 'traces over' the measuring apparatus (pointer) and/or environmental degrees of freedom, obtaining a 'reduced density matrix' for the system. But under the unitary-only assumption, the reduced density matrix is an *improper mixture*; that is, one that cannot be interpreted as an epistemic mixture in which the system has in fact achieved a particular outcome and the probabilities weighing the various outcomes are simply measures of our ignorance. The system's density matrix may be virtually diagonal, which reflects loss of interference, but for an improper mixture, elimination of interference is not equivalent to the elimination of a superposition of outcomes. Thus, the reduced density matrix does not license a conclusion that the system can be physically described by the eigenstate corresponding to the observed outcome.

---

[1] I do not attempt here to do full justice to the history of the decoherence program. Key pioneering works are: Joos, E., Zeh, H.D. (1985), Omnès, R. (1997), Zurek, W.H. (2003). Additional relevant references may be found in Kiefer C., Joos E. (1999).

Consequently, in the unitary-only (UO) approach, the use of a reduced density matrix to predict a system's measurement outcomes can be no more than a 'For All Practical Purposes" (FAPP) procedure.[2] Because the mixture UO yields is improper, it cannot explain why any system ever exhibits a determinate outcome—that is, it cannot explain why we do not see a superposition of outcomes (even at the level of the measuring device). In other words, decoherence does not solve the measurement problem. All that it can explain is why the off-diagonal elements tend to vanish, thus apparently yielding a preferred basis. However, even ignoring the fact that decoherence does not solve the measurement problem, the program's initial goal of answering the question 'Why do we observe a classical spacetime?' (the title of a pioneering paper by Joos, 1986) has arguably not been met, since the argument intended to satisfy this goal suffers from circularity (e.g., Kastner, 2014; Dugić and Jeknić-Dugić, 2012; Zanardi, 2001 ).[3]

In what follows, we will review the basics of the traditional UO treatment of decoherence. We will then consider how the measurement process, including resulting decoherence and the observed determinacy of outcomes, can be given a more satisfactory, non-circular and consistent interpretation in the Transactional Interpretation (TI), which includes physical non-unitarity with respect to a particular basis: that of conserved currents such as 4-momentum and angular momentum.

## 2. Decoherence: the basics

There are two distinct aspects to decoherence: (a) the resolution of the basic induced measurement correlation with respect to the system observable under study; and (b) the rate of repetition of the physical process constituting the measurement interaction. The first aspect describes the initial sizes of the off-diagonal elements of the system's reduced density matrix, which are governed by the decoherence function; i.e., the inner product of the pointer states corresponding to the system's observable of interest. The second describes how quickly those elements diminish with continuing repeated measurement interactions. A common example of such a measurement interaction leading to decoherence is the emission of photons. An instructive account of this process is presented in Kokorowski et al (2001). They analyze an experiment with atoms in an interferometer subject to stimulating radiation that causes them to emit photons in one of two possible 'pointer' states corresponding to each path of the interferometer.

---

[2] This term was first introduced by John S. Bell (1990) to express his dissatisfaction with existing accounts of measurement in quantum theory.

[3] Zurek does attempt to address this issue in terms of a transcendental sort of argument. He says: "As the interpretation problem does not arise in quantum theory unless interacting systems exist, we shall also feel free to assume that an environment exists when looking for a resolution." (Zurek, 2003) But in the unitary-only account, interacting quantum systems need not be anything leading to classically recognizable pointer states. In order to derive a preferred pointer basis corresponding to classicality under the unitary-only restriction, special assumptions are needed, such as a computational basis preferring classically recognizable information, initial separability of selected degrees of freedom, and the environment or apparatus having many more degrees of freedom than the designated system. Arguably, such assumptions incorporate classicality at the outset, thus making the program circular.

We will ultimately see that it is in connection with (b) that the transactional picture introduces genuine physical non-unitarity, and that is what yields a proper mixture for the resulting reduced density matrix for the system. That is, under TI the system's density matrix turns out to look exactly the same as what is obtained from just 'tracing over' the pointer degrees of freedom, but the combined system has undergone a non-unitary transformation in which the 'tracing over' corresponds precisely to that non-unitarity. The 'tracing over' in the transactional case is not just a mathematical procedure by which we ignore the pointer degrees of freedom. Instead, it is the representation of a specific physical situation created through measurement in TI: the non-unitary breaking of entanglement with respect to the degree of freedom being 'traced over.' The latter aspect will be discussed in detail in §3.

2.a Resolution of the basic measurement interaction

First, let us consider (a) above: the resolution of the basic measurement interaction, governed by the decoherence function. Here, we roughly follow the pedagogical treatment of Kiefer and Joos (1999). Recall that a measurement correlation can be induced by a suitable interaction Hamitonian between the system and a measurement apparatus, such that each system state $|n\rangle$ corresponding to an eigenvalue $n$ of the observable of interest is correlated to a 'pointer' state of the apparatus, $|\varphi_n\rangle$. Depending on the nature of the interaction between the pointer and the system, these correlated pointer states need not be mutually orthogonal, though the observed result, say some value $x$, for the pointer will always be a eigenvalue of an orthogonal pointer basis $\{x\}$ for the pointer observable. As a special case, in a sharp measurement, the states $|\varphi_n\rangle$ coincide with the eigenstates for the observed pointer outcomes. However, in general that is not the case. Since these two classes of pointer states need to be carefully distinguished, let us follow Bub's approach (Bub 1997) and call the $|\varphi_n\rangle$ 'relative' pointer states, since they are defined relative to the system's eigenstates $|n\rangle$, but may not form an orthonormal set. Thus, they are not necessarily eigenstates of any well-defined pointer observable, while the observed pointer outcomes $\{x\}$ are always eigenvalues of a pointer observable, corresponding in this case to the eigenstates $|x\rangle$.

If the system is initially in an arbitrary state $|\psi\rangle = \sum_n c_n |n\rangle$ and the apparatus in an initial ready state $|\varphi_0\rangle$, the evolution looks like:

$$|\psi\rangle|\varphi_0\rangle \rightarrow \sum_n c_n |n\rangle|\varphi_n\rangle \tag{1}$$

After the above evolution yielding the combined total state on the right hand side of (1), tracing over a pointer basis yields the reduced density matrix for the system:

$$\rho_S = \sum_{n,m} c_m^* c_n \langle \varphi_m | \varphi_n \rangle |m\rangle\langle n| \tag{2}$$

We see that the sizes of the off-diagonal elements are governed by the inner products of the relative pointer states, $\langle \varphi_m | \varphi_n \rangle$. As noted above, this quantity is the decoherence function. For $\langle \varphi_m | \varphi_n \rangle \approx 0$, the system's post-measurement density matrix is virtually diagonal. In this latter case, we have a 'sharp' measurement interaction: one with good resolution with respect to the system observable of interest. On the other hand, for sizeable $\langle \varphi_m | \varphi_n \rangle$ (non-orthogonal relative pointer states), we have a 'weak' or 'unsharp' measurement interaction, in which the pointer states do not correlate well with the system observable of interest.

Let us consider a simple example to see how this works in practice: a two-slit experiment with slits labeled A and B. The system states corresponding to 'passage through slit A' and 'passage through slit B' are $|A\rangle$ and $|B\rangle$, respectively; this forms an orthonormal basis of the relevant 2-dimension system Hilbert space. The pointer states directly corresponding to each system state are $|x_A\rangle$ and $|x_B\rangle$ respectively, and form an orthonormal basis for the 2-dimensional pointer Hilbert space. When the pointer is read, the outcome is either $x_A$ or $x_B$, corresponding to one of these pointer eigenstates. In the usual 'sharp' or 'strong measurement,' beginning with the pointer in an initial ready state $|x_0\rangle$, the evolution for each system state looks like:

$$|A\rangle|x_0\rangle \rightarrow |A\rangle|x_A\rangle;$$
$$|B\rangle|x_0\rangle \rightarrow |B\rangle|x_B\rangle \tag{3}$$

so that for some arbitrary superposition of the system states, we get

$$(c_A|A\rangle + c_B|B\rangle)|x_0\rangle \rightarrow c_A|A\rangle|x_A\rangle + c_B|B\rangle|x_B\rangle \tag{4}$$

Now suppose that instead we have a 'weak' or 'unsharp' measurement, in which the measurement evolution for each system state $|A\rangle$, $|B\rangle$ has an error component, such that the pointer has an amplitude of $e$ to give the 'wrong' answer—e.g., to yield an outcome of

$|x_B\rangle$ even though the system is $|A\rangle$. The amplitude for the correct pointer results is $d$, where $|d^2| + |e^2| = 1$. The evolution for the system states then looks like:

$$|A\rangle|\varphi_0\rangle \rightarrow |A\rangle|\varphi_A\rangle = |A\rangle(d|x_A\rangle + e|x_B\rangle);$$
$$|B\rangle|\varphi_0\rangle \rightarrow |B\rangle|\varphi_B\rangle = |B\rangle(e|x_A\rangle + d|x_B\rangle) \quad (5)$$

Thus, for an arbitrary superposition as in (4) above, we find the evolution:

$$(c_A|A\rangle + c_B|B\rangle)|\varphi_0\rangle \rightarrow c_A|A\rangle|\varphi_A\rangle + c_B|B\rangle|\varphi_B\rangle =$$
$$c_A d|A\rangle|x_A\rangle + c_A e|A\rangle|x_B\rangle + c_B e|B\rangle|x_A\rangle + c_B d|B\rangle|x_B\rangle \quad (6)$$

If we then trace over the pointer, we find for the system's reduced density matrix (where $i,j = \{A,B\}$):

$$\rho_S = \sum_{i,j} c_j^* c_i \langle \varphi_j | \varphi_i \rangle |i\rangle\langle j| = \begin{bmatrix} |c_A|^2 & c_A^* c_B \langle \varphi_A | \varphi_B \rangle \\ c_B^* c_A \langle \varphi_B | \varphi_A \rangle & |c_B|^2 \end{bmatrix} \quad (7)$$

Note that $\langle \varphi_i | \varphi_j \rangle = \langle \varphi_j | \varphi_i \rangle = d^* e + e^* d$. For vanishing error amplitude, $e=0$, we would get a sharp measurement with respect to the 'which slit' basis, and the system's reduced density matrix would be diagonal in that basis. This shows why it is the inner product of the relative pointer states (decoherence function) that specifies the resolution of the basic measurement interaction regarding the system observable of interest. An inner product of zero and a resulting diagonal reduced density matrix tells us that the possibility of a wrong pointer answer vanishes; we always get an answer that corresponds perfectly to a value of the system observable. But again, under the assumption of unitary-only evolution, this is an improper mixture, and the mathematical representation does not warrant a conclusion that any result has in fact occurred. So we have a discrepancy between the mathematical representation and what we observe, which is always a definite result. A bit later on, we will see how the transactional picture remedies this problem, but first let us consider the second aspect of decoherence: the repetition of the physical process that leads to a specific rate of decrease of the off-diagonal elements.

2b. Decay of off-diagonals due to repetition of the measurement interaction

We now consider the second aspect of decoherence discussed above, which we termed (b), involving the repetition of the measurement interaction. For a single low resolution, 'weak' or 'unsharp' measurement interaction, the system retains some coherence in that its off-diagonal elements are still of significant magnitude. However, if the same measurement interaction is repeated, the system's reduced density matrix diagonalizes rapidly, as follows. First, it is assumed that each measurement interaction is independent (this assumption becomes justified in the transactional picture, as we will see in §3c.). The process is governed by a differential equation that relates the rate of change of the off-diagonal elements $\rho_{mn}$ to a parameter $\lambda$, which is a function of the rate of repetition $\Gamma$ of the measurement interaction (such as an emission rate), as well as of the basic measurement resolution, described by the decoherence function $\langle \varphi_m | \varphi_n \rangle$. Specifically, $\lambda = \Gamma(1 - \langle \varphi_m | \varphi_n \rangle)$ and the rate of change is given by:

$$\frac{\partial \rho_{mn}}{\partial t} = -\lambda \rho_{mn} \qquad (8)$$

Integrating (8), we find that the off-diagonal elements decrease exponentially with respect to time as dictated by both the repetition rate and the resolution of the basic interaction:

$$\rho_{mn}(t) = \rho_{mn}(0) e^{-\lambda t} \qquad (9)$$

But again, under the usual unitary-only assumption, despite the vanishing of the off-diagonal elements due to this repetition, the improper mixture of the reduced density matrix does not reflect any determinacy of outcome and therefore cannot account for our observations in that regard. In the next section we will see how this shortcoming is remedied with the aid of the transactional picture.

## 3. The transactional picture completes the decoherence account of determinate measurement results and classical emergence.

3a. A brief review

The Transactional Interpretation (TI) is based on the direct-action or 'absorber' theory of fields, first introduced by Wheeler and Feynman (1945, 1949). The novel feature of the absorber theory was the idea that both emitters and absorbers generate time-symmetric fields, where absorbers respond to the field of an emitter, with the opposite phase. At the non-relativistic level, the emitter/absorber interaction is such that a net retarded field arises between the emitter and absorber, while all residual advanced effects cancel. For details, see Cramer (1986) and Kastner (2013), Chapter 3. In TI, the retarded field from the emitter is called an 'offer wave' and corresponds to the usual quantum state $|\Psi\rangle$, while the advanced field from absorbers, corresponding to the dual vector $\langle\Psi|$, is called a 'confirmation wave'. (More precisely, for N absorbers, each absorber responds with the adjoint vector of the component of $|\Psi\rangle$ received by it; more details are given in what follows.) Offer waves have all the usual features of standard quantum theory, such as entanglement, etc., depending on the specifics of the preparation.

It is absorber response that contributes a form of non-unitarity lacking in the standard quantum theory. Thus, TI involves explicit collapse or reduction, yielding a well-defined measurement result; this is discussed in Kastner and Cramer (2018), henceforth KC2018. However, TI differs significantly from the usual 'collapse theory' or 'collapse interpretation,' in that it does not make any *ad hoc* change to the basic quantum theory. Indeed, at the quantum relativistic level, it is a theorem that the 'absorber' theory of fields yields the very same matrix elements as standard quantum field theory for all cases in which (in the usual parlance) there are no 'external photons.'[4] This is proved, for example, in Akhiezer and Berestetskii (1965), p. 302, in which the direct-action form of the scattering matrix is obtained by setting the unsourced electromagnetic potential $A_\mu$ to zero.[5] Thus, when external free photon states are excluded, the quantum direct-action theory (QDAT) yields the very same probabilistic predictions as, and cannot be in conflict with, any experiments that yield results in concert with the predictions of standard quantum theory.[6] Though it may seem surprising that a 'collapse' theory could be empirically equivalent (in terms of the Born Rule) to standard quantum theory, indeed it is, as the theorem shows.

The only difference between TI and standard quantum theory, in terms of predictions, is that TI predicts non-unitary collapse (under well-defined circumstances, further discussed below) while standard quantum theory does not. It has been shown in Kastner (2018) that the quantitative circumstances of collapse under TI yield results consistent with macroscopic observations. TI could be experimentally distinguished from other 'collapse' theories, such as the GRW theory (Ghirardi-Rimini-Weber, 1986), insofar as those other theories departed from the predictions of standard quantum theory (which are upheld in TI).

---

[4] Note that it could never be empirically checked whether 'external free photons' exist. External free photon states are a theoretical construct without empirical correlate, and are therefore not an obligatory part of the physical description.

[5] The relevant result in Akhiezer and Berestetskii is their (24.25): ,
$S = T \exp\left(-\frac{1}{2} \int dx\, dy\, j_\mu(x) D_F(x-y) j^\mu(y)\right) \times N_A \exp\left(i \int dx\, j_\mu(x) A^\mu(x)\right)$, where T is a time-ordering operator and $N_A$ is a number-ordering operator. Setting $A_\mu$ to zero straightforwardly gives the direct-action form of the scattering matrix. The exclusion of external photon states is the quantum analog of the 'light tight box' condition, which is not a cosmological constraint at the quantum level but rather is just a form of the completeness condition; i.e., there must be sufficient absorber response to effect cancellation of any advanced field components to the past of the emitter and of any retarded field components to the future of the absorbers. This condition actually defines the criterion for measurement to occur, thus providing a definition of 'measurement' available in the transactional picture but lacking in standard approaches.

[6] While the usual context for application of the transactional picture involves a single emitter and several absorbers responding independently, nothing precludes higher-order processes. For example, in principle more than one absorber could 'share' a confirming response to a given OW. This requires a high degree of mutual coherence of the absorbers, and to the extent it is possible, is an exotic and short-lived effect. Such a situation is discussed in Grangier, Aspect and Vigué (1985), although what they demonstrated may be more accurately described as a 'resonance' phenomenon than an actual absorption.

Measurement in TI is well defined and based on absorber response. However, the notion of 'absorber response' is really only applicable at the classical and quantum non-relativistic level. At the fully relativistic level of RTI, 'absorber response' is really a mutual non-unitary interaction between emitter and absorbers; see also note 5). This process is discussed in detail in KC2018. The probability for such a measurement interaction can be precisely quantified in terms of emission or decay rates. So, for example, a single photon may be emitted (prepared) and absorbed (detected); that process is quantitatively well-defined in the direct-action picture and is explicitly found to be a non-unitary process constituting measurement, or more generally, objective quantum state reduction, as shown in KC2018.[7]

For present purposes, it may be recalled that in the 'direct-action' or absorber theory, charges interact by way of the time-symmetric field propagator rather than the Feynman propagator, and that the basic field is non-quantized (Davies 1971, 1972). Charges in excited states (such as an electron in an excited atomic state) are eligible to function as emitters, while charges subject to excitation satisfying 4-momentum conservation are eligible to function as absorbers for a given emitter. In general, an emitter E interacts with many eligible absorbers A. The retarded component from the emitter received by a particular absorber $A_k$ is $\langle k|\psi\rangle|k\rangle$, where $|\psi\rangle$ is the emitted state. As noted above, traditionally in TI this is referred to as an 'offer wave,' or just 'offer'. Absorber $A_k$ responds with its own time-symmetric field, the advanced component of which, $\langle\psi|k\rangle\langle k|$, proceeds back to the emitter E. This response is called a 'confirmation wave' or confirmation. The outer product of these states yields the weighted projection operator $|\langle\psi|k\rangle|^2 |k\rangle\langle k|$, where the weight factor $|\langle\psi|k\rangle|^2$ is just the Born Rule for the probability of outcome $k$ given preparation in the state $|\psi\rangle$; thus, the Born Rule gains physical justification in TI. The possible outcomes correspond to real photons of momentum $k$, i.e., Fock states, and are thus quantized forms of the field. The origin in the formalism of outer products such as $|\langle\psi|k\rangle|^2 |k\rangle\langle k|$ is made explicit in KC2018.

Quantization of the field arises from the responses of absorbers to the field of the emitter, which gives rise to an effective 'free field' (solution to the homogeneous equation) between emitter and responding absorbers. As noted above, this free field takes the form of a weighted sum of projection operators that can be expressed in terms of the usual quantized field, as shown in Davies (1972, pp. 1030-2).[8] This result clearly displays the non-unitary character of the process resulting from absorber response, since the weighted sum arising from responses of all eligible absorbers $A_k$ corresponds to the von Neumann 'measurement transition' or "Process 1" (von Neumann, 1932). In the usual approach, outer products are obtained only through the *ad hoc* procedure of 'writing down the density operator' corresponding to a particular quantum state. In the absorber theory, we see that the weighted projection operators, whose sum is the result of the von Neumann 'measurement transition,' arise naturally from a physical process. The realization of a particular outcome represented by $|k\rangle\langle k|$ is then understood as a process of spontaneous symmetry breaking, analogous to that which occurs in the Higgs mechanism.[9]

---

[7] That is, the term 'measurement' here does not imply any conscious observer or any intent to gain knowledge. It is a physical process that occurs in nature, i.e., a form of objective reduction, although of course it can be used for knowledge-gaining activities.

[8] The origin of the outer product form is in the factorizability of the free field in its vacuum expectation value representation, as in equation (19) of the given reference. Davies viewed this as a 'formal' procedure, but his analysis was undertaken before Cramer's development of the transactional interpretation (Cramer, 1986), which explicitly applied the absorber theory to measurement in quantum theory, connecting the field representation to the Born Rule. In light of the latter, Davies' treatment can be seen as the appropriate mathematical description of a specific physical process: the non-unitary creation and destruction of a real photon conveyed between an emitter and absorber.

Thus, to recap, the interaction between an emitter and a set of absorbers is just the Von Neumann "measurement transition" or "Process 1":

$$|\psi\rangle \rightarrow \sum_k |\langle k|\psi\rangle|^2 |k\rangle\langle k| \qquad (10)$$

where each weighted projection operator is an *incipient transaction* for varying values of $k$. Only one of these, with the appropriate Born probability, becomes the *actualized transaction*, corresponding to transfer of a real (on-shell) photon from the emitter to the *receiving absorber*. For an introductory discussion of this feature of TI, including the origin of the Born Rule through the inclusion of absorber response, see Kastner (2013), Chapter 3, or Kastner (2016).

The upshot of the above is that the mixed state (10) describes a physically warranted Boolean probability space. Its status as a proper mixture is justified by the non-unitary transition precipitated by absorber responses, which (together with the applicable Hamiltonian) define the eigenbasis for the observable subject to measurement. In this typical example, it is directional momentum for a particular value of energy, which serves to localize the emitted photon to a particular absorber and also yields directional information about whatever particle emitted it, thus contributing to exact (as opposed to approximate), physically grounded decoherence. In this case, unlike under the unitary-only assumption, the physics leads to genuine localization and classical determinacy, without the need to circularly presuppose a classical 'logical basis' or separable systems 'out of the starting gate': systems are naturally separated from their entanglements via the non-unitarity inherent in ubiquitous radiative processes. Because radiative processes trigger non-unitarity and reduction, the natural basis for decoherence is that of the conserved quantities, such as 4-momentum, transferred from emitters to absorbers. Position only appears to be a preferred basis because reduction commonly takes place with respect to directional momentum, which singles out one micro-absorber (such as an atom) as opposed to others; thus a particular atom becomes excited, which serves to localize the outcome at that atom. However, the reduction takes place with respect to the observable corresponding to the conserved quantity that is actually transferred.

---

[9] E.g., Higgs (1964).

Rather than being based on an *ad hoc* choice, the fundamentality of the 4-momentum (and angular momentum) basis arises from the fact that at the relativistic level, there are no well-defined spatiotemporal observables. Instead, the observables corresponding to the conserved quantities are the ones that remain physically well-defined at all levels. This asymmetry between the spatiotemporal parameters and energy/momentum observables is harmonious with the fact that relativistic quantum field theories require the 'demotion' of position to a parameter on equal footing with the time index, which is not a well-defined observable even at the non-relativistic level.

An additional observation is in order regarding the relation of emitters and absorbers to the emitted/absorbed field. The above transactional account applies to the transfer of bosonic fields, i.e., gauge fields, and primarily the photon, which is by far the most prevalent quantum transferred in transactions used in measurement-type processes. Matter fields, such as electrons and hadrons, and bound states, such as atoms, serve as sources of the gauge fields. Unlike gauge bosons, these sources are not themselves transferred via offers and confirmations in transactions. Their participation is as emitters and absorbers, i.e., the endpoints of transactions. An atom participating as a source (emitter) or sink (absorber) in a transaction involving transfer of a real photon will itself be actualized in a state corresponding to the emission of absorption of the photon. For example, an atom can be prepared in a superposition of 'which slit' states, but upon emission of a photon of momentum *k* in an actualized transaction, the atom transitions to a state corresponding to that momentum value (we will discuss this in more detail below).

In this way, quantum systems with nonvanishing rest mass, such as electrons, hadrons, and atoms, are transformed into specific quantum states indirectly through their participation in photon emission and absorption.[10] That is, a field source does not itself have to be detected/absorbed in order to undergo objective reduction to a particular quantum state. The reduction is mutual, in that field sources are themselves actualized in specific states corresponding to their emission or absorption of an actualized photon. An electromagnetic field source, such as an electron, can emit and absorb as a component of a bound state, such as an atom. As a result, the entire atom is affected; for example, by dropping to a lower energy state and changing its center-of-mass momentum as a result of an emission of a photon by the bound electron. This is one way in which entire atoms can be detected in experiments probing their behavior.[11] An electron can also absorb a photon in an ionizing interaction, be liberated from a bound state, and in turn can then emit a photon in the inverse process, radiative recombination, to become part of a new bound state. The latter type of processes are involved in experiments involving free electrons.

Finally, it should be noted that under TI all interactions, including radiative interactions, are described by the same Hamiltonians as in standard quantum theory. So for example, the emission of photons from an excited atom as in Kokorowki *et al* (2000) proceeds with exactly the same interaction Hamiltonian. The only difference is that emission and absorption result in non-unitary reduction, as shown in KC2018.[12]

---

[10] This feature also explains why the relativistic version of TI, RTI, is immune to the Maudlin contingent absorber challenge (Maudlin, 1996). For details, including a quantitative explanation of the asymmetry between the gauge fields and their sources, see Kastner (2019a).

[11] For specific examples of modes of detection of neutral atoms based on various types of interactions with the electromagnetic field, see Wilzbach *et al* (2006).

### 3b. TI yields an epistemic interpretation of decoherence functions

Now let us return to the above two-slit example, and assume that the particle subject to the two-slit apparatus is an atom that could emit a photon. As noted above, such a photon emission could constitute a measurement interaction, in that the emitted photon would be correlated to some extent with the atomic 'which slit' state. But again, in contrast to the usual approach, under TI 'measurement' is not limited to the unitary interaction but also has a non-unitary component, yielding reduction and a definite (even if unknown) result. Let us assume that the system of interest is an atom or molecule that can emit one or more photons (e.g. by way of external excitation if necessary). The photons serve as 'pointers'. The atom is prepared in an arbitrary 'both-slits' state as in (4). If the atom is allowed to emit a photon, the combined system can be represented by a state like (6),

$$(c_A|A\rangle + c_B|B\rangle)|\varphi_0\rangle \to c_A|A\rangle|\varphi_A\rangle + c_B|B\rangle|\varphi_B\rangle =$$
$$c_A d|A\rangle|k_A\rangle + c_A e|A\rangle|k_B\rangle + c_B e|B\rangle|k_A\rangle + c_B d|B\rangle|k_B\rangle \qquad (11)$$

Here, we rewrite the photon states in terms of the relevant correlated ***k*** values (since, as noted above, *x* is not really an observable and there is no real, physical position basis {*x*}). Now, under TI, emission and absorption of the photon means that the electromagnetic field undergoes a non-unitary transformation corresponding to von Neumann's Process 1, as discussed above. In addition, the atom, as an emitter, undergoes a real state change as a result of the photon's emission and absorption.

For present purposes, we suppose that there are just two photon detectors D$_A$ and D$_B$ that correspond to the atom's passage through one or the other slit, defining a two-dimensional photon momentum subspace $\{k_A, k_B\}$. It should be noted at this point that (11) is written in terms of relative states for the photon, which prioritizes the atom's detection basis. In order to take into account the effect of the photon emission on the atom, the latter being the 'measured system,' we'll need to prioritize instead the photon detection basis and define corresponding relative states for the atom.

Rewriting the state in terms of the photon basis defines unsharp relative states $|\alpha\rangle, |\beta\rangle$ for the atom:

---

[12] Specifically, the action involving the Feynman propagator is complex, and thus the associated S-matrix is non-unitary. This feature is overlooked in standard quantum field theory, but is revealed in the absorber theory. For further insight into this issue, including how it leads to decoherence, see Breuer and Petruccione (2000), pp. 40-41.

$$|\Psi\rangle = \left(c_A d|A\rangle + c_B e|B\rangle\right)|k_A\rangle + \left(c_A e|A\rangle + c_B d|B\rangle\right)|k_B\rangle \qquad (12)$$
$$\equiv a|\alpha\rangle|k_A\rangle + b|\beta\rangle|k_B\rangle$$

where

$$|\alpha\rangle = \frac{1}{a}\left(c_A d|A\rangle + c_B e|B\rangle\right)$$
$$|\beta\rangle = \frac{1}{b}\left(c_A e|A\rangle + c_B d|B\rangle\right) \qquad (13)$$

and

$$|a|^2 = |c_A d|^2 + |c_B e|^2 ;$$
$$|b|^2 = |c_A e|^2 + |c_B d|^2 \qquad (13a)$$

It may be noted that the quantities (13a) define the magnitudes *a* and *b* of the atomic state components that correspond to each of the possible photon emission/detections. That is, these components are simply $a|\alpha\rangle$ and $b|\beta\rangle$, respectively (see the first line of (12), in parentheses, for their explicit form). The magnitude of each of these atomic state components determines the amplitude of the component of the photon offer wave reaching each respective detector, with a corresponding confirmation generated. So, for example, the component of the photon offer wave having momentum **k**$_A$ is generated by atomic state component $a|\alpha\rangle$, so it has an amplitude of *a*. Thus, what reaches the photon detector D$_A$ is the photon component $a|k_A\rangle$. That prompts a confirmation $a^*\langle k_A|$, with the resulting incipient transaction being represented by the outer product $|a|^2|k_A\rangle\langle k_A|$. The weight $|a|^2$ is the probability of actualization of that incipient transaction.

An actualized photon transaction at D$_A$, actualizing the outcome $|k_A\rangle\langle k_A|$, leaves the atom definitively in the state $|\alpha\rangle$, and *mutatis mutandis* for D$_B$. At this point, the photon's quantum state has undergone reduction, so that upon absorption it conveys a determinate value of momentum **k**$_A$ or **k**$_B$, whether or not we know what that is at any given time. If we now want to describe the total system in terms of a density matrix $\rho$ when reduction has occurred, but we don't know which of the photon's states and associated atomic states has been actualized, we legitimately construct an *epistemic* mixture consisting of the incipient transactions, which are just the weighted projection operators for those possible final states:

$$\rho = |a|^2|\alpha\rangle\langle\alpha|\otimes|k_A\rangle\langle k_A| + |b|^2|\beta\rangle\langle\beta|\otimes|k_B\rangle\langle k_B| \qquad (14)$$

If we look only at the atomic subspace without specifying any detection scheme for the atom, we find its reduced density matrix:

$$\rho_S = |a|^2 |\alpha\rangle\langle\alpha| + |b|^2 |\beta\rangle\langle\beta| =$$
$$(c_A d|A\rangle + c_B e|B\rangle)(c_A^* d^* \langle A| + c_B^* e^* \langle B|) +$$
$$(c_A e|A\rangle + c_B d|B\rangle)(c_A^* e^* \langle A| + c_B^* d^* \langle B|) =$$
$$\begin{bmatrix} |c_A|^2 (d^*d + e^*e) & c_A^* c_B (d^*e + e^*d) \\ c_B^* c_A (d^*e + e^*d) & |c_B|^2 (d^*d + e^*e) \end{bmatrix} = \begin{bmatrix} |c_A|^2 & c_A^* c_B \langle\varphi_A|\varphi_B\rangle \\ c_B^* c_A \langle\varphi_B|\varphi_A\rangle & |c_B|^2 \end{bmatrix}$$ (15)

Thus, we obtain the same expression as in the standard decoherence approach for the system's reduced density matrix, but in this case, tracing with respect to the photon pointer subspace corresponds to taking into account the detection of the photon and its effect on the atom, but not yet the final detection scheme for the atom. That is, (15) describes the system when the photon it emits is really absorbed at a particular detector in a non-unitary process.[13] So we are not merely ignoring an assumed ongoing entanglement and using an improper mixture; instead, we are dealing with a physically justified proper mixture that can be interpreted epistemically. The atom *really* acquires the state $|\alpha\rangle$ or $|\beta\rangle$ corresponding to a photon detection at $D_A$ or $D_B$. so (15) is an epistemic mixture of those states. But since in general (owing to the error component) these are superpositions of the 'which slit' states, the atom's density matrix retains non-vanishing off-diagonal elements indicating retained coherence to some degree. This will be reflected in the probabilities for the atom's final detection at a screen or interferometer detector, which will provide data on interference fringe visibility.

For example, fringe visibility will be maximized for maximal error, $d=e=\frac{1}{\sqrt{2}}$, since in this case the relative atomic states coincide to the initial prepared state, i.e.: $|\alpha\rangle = |\beta\rangle = c_A|A\rangle + c_B|B\rangle$. In this limit, no entanglement of the quanta is created through the interaction. The atom emits a photon in a 'both slits' state, $|k_+\rangle$, having no dependence on the atom's prepared state; thus the total system remains a product state. That is, no entanglement is created through the photon emission. Meanwhile, at the opposite limit of zero error ($d=1, e=0,$), fringe visibility will be absent, since the relative atomic states then coincide with the orthogonal 'which slit' states: $|\alpha\rangle = |A\rangle, |\beta\rangle = |B\rangle$. It should be noted, however, that for the case in which $c_A = c_B = \frac{1}{\sqrt{2}}$, there is no preference for the 'which slit' basis even though interference is absent for the total detection distribution (i.e., without sub-ensembles obtained through coincidence counting). This is because the state can equally well be written in the 'both slits' basis, which yields exactly the same total distribution as the 'which-slit' basis. This fact is behind the phenomena observed in so-called 'quantum eraser' (QE) experiments (Kastner 2019b). (For a transactional account of the QE, see Fearn (2016) and Kastner (2013), Chapter 5.)

---

[13] Recalling the remark at the end of §3a, 'absorbed at a detector' means different physical things depending on whether the system is a gauge boson, like a photon, or a fermionic matter system, such as electron or atom. A photon is created and absorbed through direct interaction of the time-symmetric electromagnetic fields, while fermionic matter systems are detected indirectly through their coupling with photons. The latter involves secondary transactions, mediated by photon transfer, between the matter systems and their detectors. Thus, all detections involve real non-unitarity, but the process takes different forms depending on the type of system being detected.

Let us now take a closer look at the details for cases in which we explicitly observe interference effects, such as by detection of the atom at a screen. In this case the relevant detection basis for the atom is the (discrete) transverse screen coordinate $x$ corresponding to the screen's absorbing pixels. For simplicity, let us assume $c_A = c_B = \frac{1}{\sqrt{2}}$, so for the atom's relative states we have:

$$|\alpha\rangle = d|A\rangle + e|B\rangle$$
$$|\beta\rangle = e|A\rangle + d|B\rangle \tag{16}$$

And since $a = b = \frac{1}{\sqrt{2}}$, the total state, from eqn. (12), is:

$$|\Psi\rangle = \frac{1}{\sqrt{2}}\left[|\alpha\rangle|k_A\rangle + |\beta\rangle|k_B\rangle\right] \tag{17}$$

The atom's reduced density matrix $\rho_s$ is

$$\rho_s = \frac{1}{2}(|\alpha\rangle\langle\alpha| + |\beta\rangle\langle\beta|) = \begin{bmatrix} \frac{1}{2} & \frac{1}{2}\langle\varphi_A|\varphi_B\rangle \\ \frac{1}{2}\langle\varphi_B|\varphi_A\rangle & \frac{1}{2} \end{bmatrix} \tag{18}$$

Taking into account the unitary evolution of the 'which slit' states to the final screen, we express the relative states in the pixel $X$ basis:

$$|\alpha\rangle = \sum_x \left(d\langle x|A\rangle + e\langle x|B\rangle\right)|x\rangle$$
$$|\beta\rangle = \sum_x \left(e\langle x|A\rangle + d\langle x|B\rangle\right)|x\rangle \tag{19}$$

Now, recall that the atom's reduced density matrix--in this case, (18)--is a sum of the weighted projection operators for each of these relative states. This corresponds (in RTI) to the fact that for each run of the experiment, the photon is detected at either $D_A$ or $D_B$ and the atom really projected into the corresponding relative state, so that there is a fact of the matter about the atom's possession of that state. Thus, the weights (which are the Born probabilities) are just measures of our ignorance about which state the atom acquires for any particular (unknown) photon detection. However, there are still transactional opportunities for the atom corresponding to the incipient transactions for each pixel *x*.

For the atomic state $\alpha$, corresponding to photon detection at $D_A$, a particular pixel *x* receives the component

$$\left(d\langle x|A\rangle + e\langle x|B\rangle\right)|x\rangle \tag{20}$$

and responds with the adjoint confirmation

$$\left(d^*\langle A|x\rangle + e^*\langle B|x\rangle\right)\langle x| \tag{21}$$

and the incipient transaction is described by their outer product,

$$\left|\left(d\langle x|A\rangle + e\langle x|B\rangle\right)\right|^2 |x\rangle\langle x| \tag{22}$$

Owing to the squared sum of amplitudes for nonvanishing error *e*, the probability distribution exhibits the usual interference. Analogous incipient transactions, for each value of x, arise for the atomic state $\beta$. corresponding to photon detector $D_B$. Thus, the observed total probability distribution *P(x)* on the screen will be the sum

$$P(x) = \frac{1}{2}\left\{\left|\left(d\langle x|A\rangle + e\langle x|B\rangle\right)\right|^2 + \left|\left(e\langle x|A\rangle + d\langle x|B\rangle\right)\right|^2\right\} \tag{23}$$

where the two terms correspond to atomic states $\alpha$ and $\beta$ for photon detections at $D_A$ and $D_B$ respectively. From this, we can note that for a sharp measurement (d=1, e=0), we just get the usual sum of two independent 'which slit' states. In contrast, for maximal error, $d = e = \frac{1}{\sqrt{2}}$, we get maximal fringe visibility due to the fact that the states $\alpha$ and $\beta$ coincide.

Of course, the probabilities are the same as in the standard account, but in the transactional picture we gain a physical reason for the squaring procedure of the Born Rule, as well as a physical basis for the interpretation of the system's reduced density matrix as a legitimate epistemic mixture. Measurements really do have results, and they have results in virtue of the advanced responses of absorbers, which are missing in the standard account.

Thus far, we have considered a single measurement process which, in the case of nonvanishing error *e* for the which-slit correlation, has nonzero off-diagonal entries for the system's reduced density matrix $\rho_S$. The derived $\rho_S$ yields the same 'decoherence function' $\langle \varphi_m | \varphi_n \rangle$ for the off-diagonal elements that is obtained by tracing over the pointer degree of freedom in the standard unitary-only approach. However, it is a proper mixture in the transactional picture, because measurements really are non-unitary processes that lead to determinate results.

3c. Vanishing of interference effects with real quantum state reduction

We now return to the second aspect of decoherence, the vanishing of the off-diagonals with repeated measurements. From this point, the matter is trivial, since as we have seen above, a measurement in the transactional picture involves non-unitary reduction of the quantum state and the legitimate interpretation of the reduced density operator as reflecting an epistemic mixed state for the system. All we need do is to note that repeating the non-unitary measurement process is subject to the same analysis as in Section 2b; i.e., equation (8) applies. Besides the benefit gained in applying (8) to what we now know is a proper mixture, an additional dividend of the transactional picture is that we do not need to assume (without obvious justification) that the measurement events are independent. In fact, they are shown to be independent through the relativistic development of the transactional picture, RTI, which derives from basic physical principles of the direct-action theory the fact that measurement interactions obey Poissonian statistics corresponding to standard decay rates (Kastner and Cramer 2018). Thus, the current approach provides a rigorous account of the onset of Markov behavior.

The reductions producing the specific outcomes are inherently unpredictable in each case; their occurrence is understood as a form of spontaneous symmetry breaking. The fact that collapse is unpredictable follows from the objective indeterminism attending quantum theory if neither hidden variables nor other sub-quantum structures are invoked. The present work takes no position on whether there are such sub-quantum structures that could lead to an account of the realization of one outcome out of many. Were that the case, it might lead to a greater understanding of other cases in physics currently attributed to spontaneous symmetry breaking (such as the 'choice' of one vacuum state out of many in the Higgs mechanism).

4. Recoherence still possible

The foregoing analysis does not in any way preclude the ability to 'recohere' a given degree of freedom, as predicted and demonstrated for example in Bouchard *et al* (2015). These authors note that a suitable inverse unitary operation applied to an entangled photon pair reverses their entanglement. This analysis pertains just as well to the current treatment, which takes into account any unitary evolution existing ahead of a final absorption opportunity. If such an inverse unitary process is applied to the entangled degrees of freedom ahead of their exposure to absorption, then their individual detectors respond to the original pure states generated prior to entanglement. Thus, recoherence is always possible if absorption is prevented prior to any application of the inverse unitary transformation.

Another way to understand the issue is in terms of Markovian vs. non-Markovian processes. The possibility of recoherence exists only in a non-Markovian process, which is characterized by the preservation of unitarity. However, according to the present formulation, the more opportunities for absorption that are present, the more difficult this becomes. In contrast, a Markovian process is describable by a master or Lindblad equation, and at that point, recoherence is no longer possible. However, Markovian processes strictly arise only in the presence of non-unitarity. The usual unitary-only assumption accompanying quantum theory precludes a rigorous, non-circular account of the onset of Markov behavior (see, e.g., Kastner 2017). The account presented herein provides a way to understand the observed emergence of Markov behavior while still allowing for recoherence at the unitary level.

5. Conclusion.

We have analyzed the process of decoherence in the relativistic Transactional Interpretation (RTI), which incorporates non-unitary quantum state reduction under well-defined physical circumstances. The account is possible because the field behavior is as described by the direct-action (absorber) theory, which includes physical non-unitarity not existing in the standard account. Non-unitarity occurs due to the response of absorbers, which can be precisely quantified (to within the uncertainty principle), as shown in Kastner and Cramer (2018).  We recover the usual empirically corroborated decoherence functions in this account, along with a physical justification for the epistemic interpretation of the reduced density matrix of measured systems.

Acknowledgments. I would like to thank two anonymous reviewers for suggestions for improvement of the presentation.